\documentclass[a4paper, preprint]{article}

\usepackage{amsmath}
\usepackage{amssymb}
\usepackage{epsfig}
\usepackage{graphicx}

\begin{document}
\begin{center}
\Large{Damped Quantum Interference using Stochastic Calculus}
\end{center}

\begin{center}
\large{D. Salgado \& J.L. S\'{a}nchez-G\'{o}mez}\\
Dpto. F\'{\i}sica Te\'{o}rica, Univ. Aut\'{o}noma de Madrid\\
28049 Cantoblanco, Madrid\\
Spain
\end{center}

\begin{abstract}
It is shown how the phase-damping master equation, either in Markovian and nonMarkovian regimes, can be obtained as an averaged random unitary evolution. This, apart from offering a common mathematical setup for both regimes, enables us to solve this equation in a straightforward manner just by solving the Schr\"{o}dinger equation and taking the stochastic expectation value of its solutions after an adequate modification. Using the linear entropy as a figure of merit (basically the loss of quantum coherence) the distinction of four kinds of environments is suggested.  
\end{abstract}

\section{Introduction}
The principle of quantum superposition is at the base both of important
issues concerning the conceptual foundations of quantum mechanics
\cite{dEsp76a} and of potential direct technological applications of
quantum theory \cite{LoPopSpi98a,BouEkeZei00a}. In particular the
preservation and dissapearance of these quantum superpositions play a
central role in topics like the quantum-to-classical transition
\cite{GiuJooKieKupStaZeh96a} and the performance of some quantum information processing
tasks \cite{ChuNie00a}. Physically this decoherence has its origins in the
pervasive effect of the surrounding environment. Mathematically it is
expressed by means of master equations (ME hereafter), that is evolution
equations for
the reduced density operator of the system which are usually
obtained after some careful assumptions \cite{CohTanDupRocGry92a}.

\bigskip

By and large two main approaches can be followed to study these master
equations \cite{Ali02a}, namely a constructive and an axiomatic
approach. In the former one begins considering the joint
system-environment degrees of freedom and finally pursues reasonable and
manageable approximations to obtain the master equation of the
system. This approach usually falls into mathematical difficulties hard to
deal with, though some limit situations have been satisfactorily dealt
with \cite{Dav76a,GorFriVerKosSud78a}. On the other hand, the latter
approach poses a set of reasonable and physically motivated assumptions
\cite{Lin76a,GorKosSud76a} and then derive the general form for the master
equation of systems satisfying that set of axioms. Though most of ME of
practical relevance fall into this category (see e.g. \cite{IsaSanSch93a}), the range of
applicability of this approach is limited by definition, especially
regarding one of its fundamental axioms, namely the semigroup condition
(Markovianity).

\bigskip

In an attempt to embrace also nonMarkovian situations we have combined
well-known stochastic methods jointly with
common tools of functional analysis employed in quantum mechanics
\cite{Pru81a,DunSchw63a} to state the following result:
\emph{Any Lindblad-type evolution with selfadjoint Lindblad operators,
\textbf{whether Markovian or nonMarkovian}, can be understood as an
averaged random unitary evolution}. This result, whose fully-fledged
mathematical proof is provided elsewhere \cite{SalSan02b},
reveals the fact that the Lindblad structure of ME for the case of
selfadjoint Lindblad operators is beyond the Markov approximation and can
also be recovered in nonMarkovian situations.

\bigskip

Here we exploit the mathematical advantages of this approach to study the
loss of coherence described by the phase-damping ME and identify different
types of environments using the linear entropy as a figure of merit.

\section{The Phase-Damping Master Equation as an Averaged Random
Evolution}

The generalized phase-damping ME reads

\begin{equation}\label{PDME}
\frac{d\rho(t)}{dt}=-i[H,\rho(t)]-\frac{\Dot{\lambda}(t)}{2}[H,[H,\rho(t)]]
\end{equation}

\noindent where $H$ denotes the Hamiltonian of the system and we have allowed
$\dot{\lambda}(t)$ to be any nonnegative real function. In general
eq. \eqref{PDME} denotes a nonMarkovian evolution reducing to a Markovian
one only when
$\Dot{\lambda}(t)=\lambda\equiv\textrm{const}$. Eq. \eqref{PDME} is a
Lindblad-type ME with selfadjoint Lindblad operators and consequently
following the result stated above can be obtained as an averaged random
unitary evolution. The proof for this case is straighforward. Let
$U(t)=e^{-itH}$ $(\hbar=1)$ be the unitary evolution operator of the
quantum system under study. Let us now define the random unitary evolution
operator
$U_{st}(t)=e^{-itH-iH\int_{0}^{t}\sigma(s)d\mathcal{B}_{s}}$, where
$\sigma(t)$ denotes an arbitrary real function and $\mathcal{B}_{t}$
denotes standard real Brownian motion \cite{Oksendal98a}, thus
$\int_{0}^{t}\sigma(s)d\mathcal{B}_{s}$ will denote the well-known
Ito integration. Now let the density operator $\rho(t)$ be defined as

\begin{equation}
\rho(t)=\mathbb{E}[U_{st}(t)\rho(0)U_{st}^{\dagger}(t)]
\end{equation}

\noindent where $\mathbb{E}$ denotes expectation value with respect to the
probability measure of $\mathcal{B}_{t}$. Then after making some
calculations one can see that $\rho(t)$ satisfies the ME \eqref{PDME}
 with $\Dot{\lambda}(t)=\sigma^{2}(t)$.

\section{Interference Damping for Plane Waves}

One of the immediate consequences of the above procedure arises by 
realizing that the solutions to the ME \eqref{PDME} can be obtained by 
performing the substitution $t\to 
t+\int_{0}^{t}\sigma(s)d\mathcal{B}_{s}$ in the unitary solutions 
and then taking the stochastic expectation value.

\bigskip

We will illustrate these ideas with the study of the interference pattern 
between two equally-weighted plane waves with momentum $k_{1}$ and $k_{2}$ 
respectively evolving under the ME \eqref{PDME}. The unitary solution can 
be obtained elementarily either by solving the Schr\"{o}dinger equation 
for the state vector or by solving the Liouville-von Neumann equation for 
the density operator or by resorting to the Wigner characteristic function \cite{Lou73a}
$\chi(\lambda,\mu)=\textrm{Tr}[\rho(t)e^{i(\mu p+\lambda q)}]$ where $q$ 
and $p$ are the position and momentum operators (cf. also \cite{SavWal85b}). The interference pattern 
is provided by the well-known expression

\begin{equation}
I(x,t)=1+\cos[(k_{1}-k_{2})x-\frac{k_{1}^{2}-k_{2}^{2}}{2m}t]
\end{equation}

This pattern is changed when the initial superposition evolves under eq. 
\eqref{PDME} and accordingly the solution is more involved. One can again 
resort to the Wigner characteristic function $\chi(\lambda,\mu)$ to obtain 
the solution \cite{SavWal85b}, but using our previous result one can 
immediately arrive at the desired expression:

\begin{equation}
I(x,t)\to\bar{I}(x,t)=\mathbb{E}[I(x,t+\int_{0}^{t}\sigma(s)
d\mathcal{B}_{s})]
\end{equation}

Then knowing that (cf. appendix)

\begin{subequations}
\begin{eqnarray}
\mathbb{E}\left[\cos\left[\frac{k_{1}^{2}-k_{2}^{2}}{2m}\left(t+\int_{0}^{t}
\sigma(s)d\mathcal{B}_{s}\right)\right]\right]&=&e^{-\frac{1}{2}
\left(\frac{k_{1}^{2}-k_{2}^{2}}{2m}\right)^{2}
\int_{0}^{t}\sigma^{2}(s)ds}\cos\left[\frac{k_{1}^{2}-k_{2}^{2}}{2m}t\right]\nonumber\\
&&\label{ExpValCos}\\
\mathbb{E}\left[\sin\left[\frac{k_{1}^{2}-k_{2}^{2}}{2m}\left(t+\int_{0}^{t}\sigma(s)d\mathcal{B}_{s}\right)\right]\right]&=&e^{-\frac{1}{2}
\left(\frac{k_{1}^{2}-k_{2}^{2}}{2m}\right)^{2}
\int_{0}^{t}\sigma^{2}(s)ds}\sin\left[\frac{k_{1}^{2}-k_{2}^{2}}{2m}t\right]\nonumber\\
\label{ExpValSin}&&
\end{eqnarray}
\end{subequations}

we get

\begin{equation}
\bar{I}(x,t)=1+e^{-\frac{1}{2}
\left(\frac{k_{1}^{2}-k_{2}^{2}}{2m}\right)^{2}
\int_{0}^{t}\sigma^{2}(s)ds}\cos[(k_{1}-k_{2})x-\frac{k_{1}^{2}-k_{2}^{2}}{2m}t]
\end{equation}

Different interference patterns are obtained corresponding to different types of environments. Note that since the Markovian approximation is not assumed, we have now gained some generality.

\section{Different Types of Environment}
We will investigate the time evolution of the linear entropy in order to 
study the progressive loss of coherence of systems driven by the ME 
\eqref{PDME}. To avoid mathematical nuisances because of the lack of 
normalizability of plane waves, let us focus upon an initial gaussian wave 
packet 
$$\psi(x,0)=\frac{1}{(2\pi\sigma^{2}_ {0})^{1/4}}e^{-\frac{x^{2}}{4\sigma^{2}_{0}}}$$ 

Under equation \eqref{PDME} the linear entropy $S_{lin}(t)$ of the system 
can be calculated:

\begin{equation}\label{LinEnt}
S_{lin}(t)=1-\frac{1}{\sqrt{16\pi}}\frac{4m\sigma_{0}^{2}}{\lambda^{1/2}(t)}\int_{0}^{\infty}\frac{dy}{y}e^{-\frac{4m\sigma_{0}^{2}}{\lambda^{1/2}(t)}y}\textrm{erf}(y)dy
\end{equation}

\noindent where $\lambda(t)=\int_{0}^{t}\sigma^{2}(s)ds$. Note that the previous integral exists for all possible choices of $\sigma(t)$ since the integrand ($\equiv g(y,t)$) is continuous in $\mathbb{R}^{+}$ and 

\begin{eqnarray}
\lim_{y\to 0}g(y,t)&=&\frac{2}{\sqrt{\pi}}\nonumber\\
\lim_{y\to\infty}g(y,t)&=&0\nonumber\\
g(y,t)&\sim&\frac{e^{-\frac{4m\sigma_{0}^{2}}{\lambda^{1/2}(t)}y}}{y}\quad y\gg 1 \nonumber
\end{eqnarray}

\bigskip

Also notice that the second term in \eqref{LinEnt} $(\equiv I(t))$ is bounded from above and from below:

\begin{equation}
0\leq I(t)\leq 1\Rightarrow 0\leq S_{lin}(t)\leq 1
\end{equation}

In particular $S_{lin}(t)=0$ if and only if $\lambda(t)=0$, i.e. one always obtains a decohering process. Immediately one can distinguish four different kinds of situations:

\begin{itemize}
\item[a)] $\lambda(t)=\lambda t$. This is the usual Markovian approximation. 
In this case (cf. also \cite{GiuJooKieKupStaZeh96a})

\begin{equation}
S_{lin}(t)\stackrel{t\to\infty}{\longrightarrow}1
\end{equation}
\item[b)] $\lambda(t)\stackrel{t\to\infty}{\longrightarrow}\lambda_{0}$. Let us call it SubMarkovian regime. 
Now the linear entropy reaches a maximum value $S_{0}<1$, thus the 
initial coherence is partially preserved.

\item[c)]$\lambda(t)\stackrel{t\to\infty}{\longrightarrow}\infty$ but $\frac{\lambda(t)}{t}\stackrel{t\to\infty}{\longrightarrow}0$. Let us 
call it type-I SuperMarkovian regime. Now as in the Markovian case 
$S_{lin}(t)\to 1$ but at a slower pace. This allows the system to preserve 
a higher degree of coherence at a given time.

\item[d)]$\lambda(t)\stackrel{t\to\infty}{\longrightarrow}\infty$ and $\frac{\lambda(t)}{t}\stackrel{t\to\infty}{\longrightarrow}\infty$. Let us call 
it type-II SuperMarkovian regime. This is the most destructive case since 
$S_{lin}(t)\to 1$, that is coherence is completely lost and even faster than in 
the usual Markovian case. 

 \end{itemize}

\begin{figure}[ht]
\begin{center}
\epsfig{file=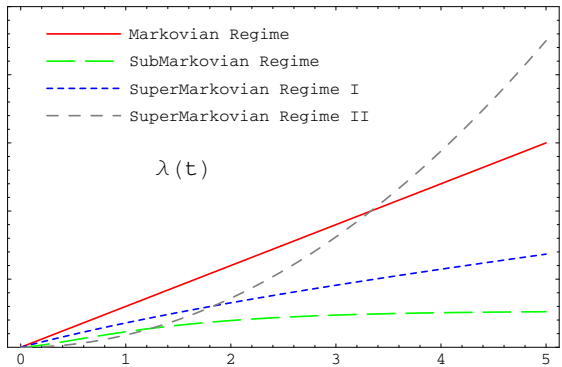,width=5cm, angle=-90}
\caption{\label{Envir}Decoherence functions for different types of environment. See text.}
\end{center}
\end{figure}

The four situations are qualitatively represented in Fig. \ref{Envir}.

\section{Conclusions}
We have derived the phase-damping master equation using stochastic 
calculus and subsequently solve it using the advantages of this procedure. 
These advantages run from embracing under the same mathematical 
formalism  both Markovian and nonMarkovian evolutions (thus allowing us to 
deal with more 
general physical situations) to providing a new method to solve this 
equation. The great advantage of this procedure comes out from the fact 
that you only need to solve the \emph{unitary} evolution equation, which 
can be done for the state vector instead of dealing with the density operator. This 
represents a computational advantage. Once the unitary solution is known 
all one has to do is to find the stochastic expectation value of this 
solution after modifying it with the appropiate random factor (see 
above).\\

Using then the linear entropy as a figure of merit we have found four 
types of environments according to the rate of loss of coherence they 
induce upon the system. In particular it may be concluded that what we 
have called SubMarkovian and Type-I SuperMarkovian regimes appear as 
better choices to perform quantum-informational and 
quantum-computational tasks than the usual Markovian environment. Now the 
physical conditions should be provided to properly identify what these 
kinds of environment are.\\ 

It should be noticed that as a by-product of this approach, independently 
of the function $\sigma(t)$ used to describe the system-environment 
response, a positive and nondecreasing rate of 
decoherence is always obtained, which indicates that it is an \emph{irreversible} process. 
Thus under the conditions in which the phase-damping master equation is 
valid to describe the evolution of a system, coherence can never be 
recovered from the environment, whether this is Markovian or 
nonMarkovian.\\

\appendix

\section{Some Computational Notes}

In order to compute the expectation values \eqref{ExpValCos} and 
\eqref{ExpValSin} it is necesarry to know the nth order moments of the 
real-valued stochastic process 
$X_{t}=t+\int_{0}^{t}\sigma(s)d\mathcal{B}_{s}$. It is simple to 
find $\beta_{n}(t)\equiv\mathbb{E}[X_{t}^{n}]$ using Ito's formula 
\cite{Oksendal98a} upon $\phi(X_{t})=X_{t}^{n}$:

\begin{eqnarray}
d\phi(X_{t})&=&(nX_{t}^{n-1}+\frac{n(n-1)}{2}X_{t}^{n-2}\sigma^{2}(t))dt+nX_{t}^{n-1}\sigma(t)d\mathcal{B}_{t}\Rightarrow\nonumber\\
\Rightarrow\frac{{d\beta}_{n}(t)}{dt}&=&n\beta_{n-1}(t)+\frac{n(n-1)}{2}\beta_{n-2}(t)\sigma^{2}(t)\quad 
n\geq 2
\end{eqnarray}
 
Since $\beta_{0}(t)=1$ and $\beta_{1}(t)=0$ it is straighforward to 
recursively arrive at

\begin{subequations}
\begin{eqnarray}
\beta_{2n}(t)&=&\frac{(2n)!}{2^{n}n!}\lambda^{n}(t)\\
\beta_{2n+1}(t)&=&0
\end{eqnarray}
\end{subequations}

\noindent where $\lambda(t)\equiv\int_{0}^{t}\sigma^{2}(s)ds$.

\bigskip

To arrive at \eqref{LinEnt} the following integral is needed which can be obtained by means of already tabulated integrals \cite{GraRiz94a}:

\begin{equation}
\int_{0}^{\pi}e^{-z\cos^{2}\theta}d\theta=\sqrt{\frac{\pi}{z}}\textrm{erf}(\sqrt{z})
\end{equation}

\section*{Acknowledgements}

One of us (D.S.) acknowledges financial support from Madrid Education Council through grant no. BOCAM-20/08/99.


\begin{thebibliography}{10}

\bibitem{dEsp76a}
B.~d'Espagnat.
\newblock {\em Conceptual Foundations of Quantum Mechanics}. 2nd compl. rev. and enl. ed.
\newblock Addison-Wesley, New York, 1976.

\bibitem{LoPopSpi98a}
H.-K. Lo, S.~Popescu, and T.~Spiller, editors.
\newblock {\em Introduction to Quantum Computation and Information}.
\newblock World Scientific, Singapore, 1998.

\bibitem{BouEkeZei00a}
D.~Bouwmeester, A.~Ekert, and A.~Zeilinger (eds.).
\newblock {\em The Physics of Quantum Information}.
\newblock Springer, Berlin, 2000.

\bibitem{GiuJooKieKupStaZeh96a}
D.~Giulini, E.~Joos, C.~Kiefer, J.~Kupsch, I.-O. Stamatescu, and H.D. Zeh.
\newblock {\em Decoherence and the appearance of a classical world in quantum
  theory}.
\newblock Springer-Verlag, Berlin, 1996.

\bibitem{ChuNie00a}
I.L. Chuang and M.A. Nielsen.
\newblock {\em Quantum Computation and Quantum Information}.
\newblock Cambridge University Press, Cambridge, 2000.

\bibitem{CohTanDupRocGry92a}
C. Cohen-Tannoudji, J. Dupont-Roc and G. Grynberg.
\newblock {\em Atom-Photon Interactions. Basic Processes and Applications}.
\newblock Wiley Interscience, New York, 1992.

\bibitem{Ali02a}
R. Alicki.
\newblock {\em Lecture given at the 38th Winter School of Theoretical Physics,
  Ladek, Poland}, 2002.

\bibitem{Dav76a}
E.B. Davies.
\newblock {\em Quantum Theory of Open Systems}.
\newblock Academic Press, London, 1976.

\bibitem{GorFriVerKosSud78a}
V.~Gorini, A.~Frigerio, M.~Verri, A.~Kossakowski, and E.C.G. Sudarshan.
\newblock {\em Rep. Math. Phys.} \textbf{13}, 149 (1978).

\bibitem{Lin76a}
G.~Lindblad.
\newblock {\em Commun. Math. Phys.} \textbf{48}, 119 (1976).

\bibitem{GorKosSud76a}
V.~Gorini, A.~Kossakowski, and E.C.G. Sudarshan.
\newblock {\em J. Math. Phys.} \textbf{17}, 821 (1976).

\bibitem{IsaSanSch93a}
A. Isar, A. Sandulescu and W. Scheid.
\newblock {\em J. Math. Phys.} \textbf{34}, 3887 (1993).
\bibitem{Pru81a}
E. Prugovecki.
\newblock {\em Quantum Mechanics in Hilbet Space}.
\newblock Academic Press, New York, 2nd edition, 1981.

\bibitem{DunSchw63a}
N.T. Dunford and J.T. Schwartz.
\newblock {\em Linear Operators. Part II: Spectral Theory}.
\newblock Interscience, New York, 1963.

\bibitem{SalSan02b}
D.~Salgado and J.L. S\'{a}nchez-G\'{o}mez,
\newblock quant-ph/0208175.

\bibitem{Oksendal98a}
B. Oksendal.
\newblock {\em Stochastic Differential Equations}. 5th ed.
\newblock Springer, Berlin, 1998.

\bibitem{Lou73a}
W.H. Louisell.
\newblock {\em Quantum Statistical Properties of Radiation}.
\newblock Wiley, New York, 1973.

\bibitem{SavWal85b}
C.M. Savage and D.F. Walls.
\newblock {\em Phys. Rev. A}, \textbf{32}, 3487 (1985).

\bibitem{GraRiz94a}
I.S. Gradshteyn and I.M. Rizhik.
\newblock {\em Table of Integrals, Series and Products}. 5h ed.
\newblock Academic Press, London, 1994.

\end{thebibliography}

\end{document}